\documentclass[preprint,pteplogo]{ptephy}

\preprintnumber{XXXX-XXXX} 

\usepackage{lscape}

\begin{document}

\title{
First demonstration of emulsion multi-stage shifter for accelerator neutrino experiment in J-PARC T60
}

\author{
\name{K. Yamada}{1},
\name{S. Aoki}{1},
\name{S. Cao}{2},
\name{N. Chikuma}{3},
\name{T. Fukuda}{4},
\name{Y. Fukuzawa}{4},
\name{M. Gonin}{5},
\name{T. Hayashino}{6},
\name{Y. Hayato}{7},
\name{A. Hiramoto}{6},
\name{F. Hosomi}{3},
\name{K. Ishiguro}{4},
\name{S. Iori}{8},
\name{T. Inoh}{8},
\name{H. Kawahara}{4},
\name{H. Kim}{9},
\name{N. Kitagawa}{4},
\name{T. Koga}{3},
\name{R. Komatani}{4},
\name{M. Komatsu}{4},
\name{A. Matsushita}{4},
\name{S. Mikado}{10},
\name{A. Minamino}{11},
\name{H. Mizusawa}{8},
\name{K. Morishima}{4},
\name{T. Matsuo}{8},
\name{T. Matsumoto}{8},
\name{Y. Morimoto}{8},
\name{M. Morishita}{4},
\name{K. Nakamura}{6},
\name{M. Nakamura}{4},
\name{Y. Nakamura}{4},
\name{N. Naganawa}{4},
\name{T. Nakano}{4},
\name{T. Nakaya}{6},
\name{Y. Nakatsuka}{4},
\name{A. Nishio}{4},
\name{S. Ogawa}{8},
\name{H. Oshima}{8},
\name{B. Quilain}{6},
\name{H. Rokujo}{4},
\name{O. Sato}{4},
\name{Y. Seiya}{9},
\name{H. Shibuya}{8},
\name{T. Shiraishi}{4},
\name{Y. Suzuki}{4},
\name{S. Tada}{4},
\name{S. Takahashi}{1,\ast},
\name{M. Yoshimoto}{4} and 
\name{M. Yokoyama}{3}
}

\address{
\affil{1}{Kobe University, Kobe 657-8501, Japan}
\affil{2}{High Energy Accelerator Research Organization (KEK), Tsukuba 305-0801, Japan}
\affil{3}{University of Tokyo, Tokyo 113-0033, Japan}
\affil{4}{Nagoya University, Nagoya 464-8601, Japan}
\affil{5}{Ecole Polytechnique, IN2P3-CNRS, Laboratoire Leprince-Ringuet, Palaiseau, France
}
\affil{6}{Kyoto University, Kyoto 606-8502, Japan}
\affil{7}{University of Tokyo, ICRR, Kamioka 506-1205, Japan}
\affil{8}{Toho University, Funabashi 274-8510, Japan}
\affil{9}{Osaka City University, Osaka 558-8585, Japan}
\affil{10}{Nihon University, Narashino 274-8501, Japan}
\affil{11}{Yokohama National University, Yokohama 240-0067, Japan}
\email{satoru@radix.h.kobe-u.ac.jp}}

\begin{abstract}%
We describe the first ever implementation of an emulsion multi-stage shifter in an accelerator neutrino experiment. The system was installed in the neutrino monitor building in J-PARC as a part of a test experiment T60 and stable operation was maintained for a total of 126.6 days. By applying time information to emulsion films, various results were obtained. Time resolutions of 5.3 to 14.7 s were evaluated in an operation spanning 46.9 days (time resolved numbers of 3.8--1.4$\times10^{5}$). By using timing and spatial information, a reconstruction of coincident events that consisted of high multiplicity events and vertex events, including neutrino events was performed. Emulsion events were matched to events observed by INGRID, one of near detectors of the T2K experiment, with high reliability (98.5\%) and hybrid analysis was established via use of the multi-stage shifter. The results demonstrate that the multi-stage shifter is feasible for use in neutrino experiments. \end{abstract}

\subjectindex{xxxx, xxx}

\maketitle


\section{Introduction}
Neutrinos have the only experimental evidence beyond the standard model, and a more thorough understanding of neutrino properties is important if we are to probe beyond our current understanding.

Nuclear emulsions, which are powerful tracking devices that can record the three-dimensional (3D) trajectory of a charged particle within a 1-$\mu$m accuracy, have already been used to perform a number of prominent observations such as the discovery of the $\pi$ meson \cite{Powell}, the discovery of the charmed particle \cite{Niu}, the first observation of tau-neutrino interactions \cite{NuTau} and the discovery of $\nu_\mu\rightarrow\nu_\tau$ appearance \cite{OPERA_NuTau}.

In recent years, significant progress has been achieved in advancing emulsion techniques. Emulsion films now provide highly precise, completely uniform, refreshable, and mass-producible tracking devices \cite{OPERAfilm}, and their development and production is now feasible at the laboratory level \cite{NewGel}. Emulsion scanning is now automated and achievable scanning speeds have been increasing exponentially\cite{TS}. Additionally, multi-stage shifter techniques now allow track-timing resolution below $\sim$seconds with high reliability and high efficiency for large-scale and inaccessible emulsion experiments \cite{MultiStageShifter}. Therefore, it is clear that the latest emulsion techniques can open the way to innovative accelerator neutrino experiments with nuclear emulsions. 

Among the studies currently being conducted at the Japan Proton Accelerator Research Complex (J-PARC) are T60 \cite{T60} experiments using the emulsion detector for feasibility studies on precise measurement of $\nu$--N interactions and strict validation of neutrino oscillation anomalies \cite{LSND, MiniBooNE}. An intense muon neutrino beam is produced by using the 30 GeV proton synchrotron at J-PARC in Tokai. The neutrino beam has a peak energy at $\sim$1 GeV on the on--axis and a cycle with 2 to 3 seconds. Specifically, a feasibility test was performed with a $\sim$kg scale target and $\sim$month scale exposure at the front of an INGRID module that had been positioned near the center of the module. The INGRID, which is an on--axis detector composed of an array of iron/scintillator sandwiches, measures the neutrino beam direction and profile for T2K experiment \cite{T2K}. Using INGRID, we implemented the first ever use of a multi-stage shifter as an emulsion detector time stamper in an accelerator neutrino experiment. The multi-stage shifter has two or more emulsion films on stages shifted with individual cycles. The combination of position displacement of the track between each stages can be taken and a many independent time-dependent states are obtained. As a result, it has good resolution for longer time periods. By combining timing information, exceptionally clear full-event reconstruction can be performed \cite{GRAINE2011}. This means that events such as neutral current single $\pi^{0}$ (1--$\pi^{0}$) production can be reconstructed via emulsion detection. It is difficult to reconstruct a 1--$\pi^{0}$ production event without timing information. Event matching between emulsion detector and INGRID can provide particle identification, especially for muons. In this paper, the first implementation and demonstration of a multi-stage shifter used for accelerator neutrino experiments is described.

\section{Setup, operation and beam exposure}
\begin{figure}[h!!!!!!]
\centering\includegraphics[width=5.0in]{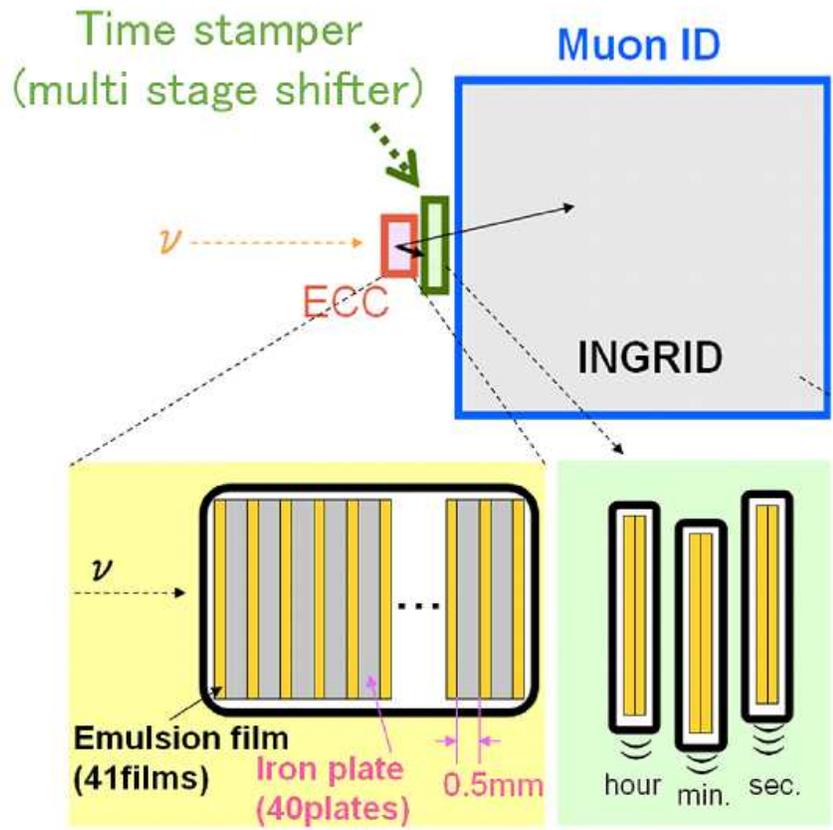}
\caption{Schematic view of experimental setup.}
\label{schematic view}
\end{figure}

\begin{figure}[h!!!!!!]
\centering\includegraphics[width=5.0in]{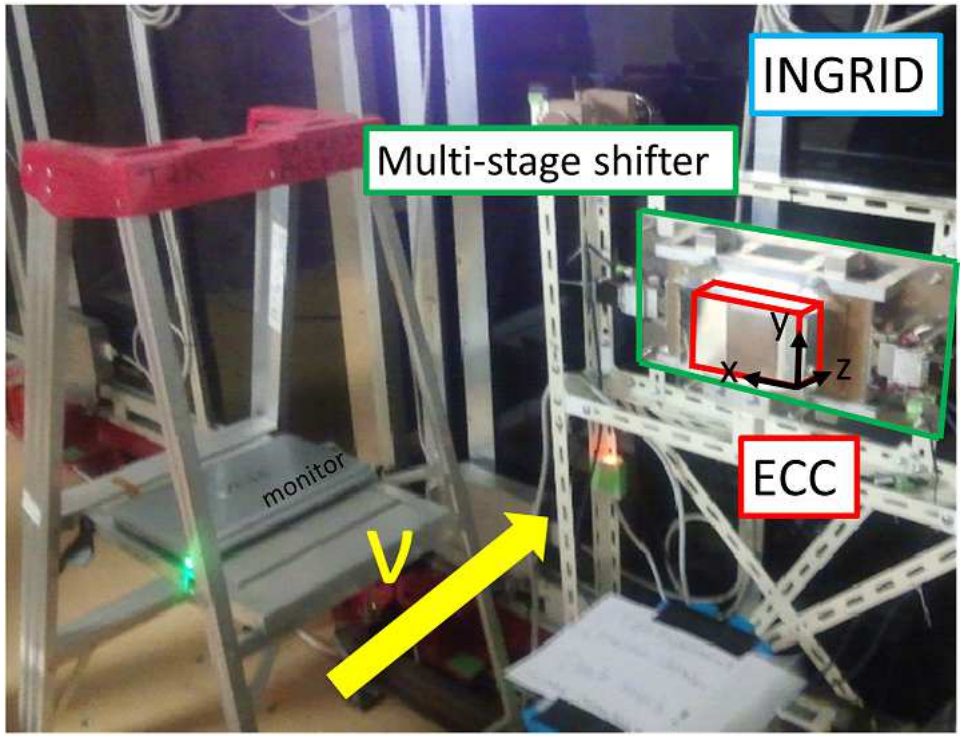}
\caption{Photograph of experimental setup. Arrows located at right--bottom of ECC show detector coordinates.}
\label{picture of setup view}
\end{figure}

Figures \ref{schematic view} and \ref{picture of setup view} show a schematic layout and photograph of the setup used in this experiment, respectively. The emulsion detector consists of an emulsion cloud chamber (ECC) for neutrino targets and a multi-stage shifter for the time stamper. The ECC has a 12.5 cm$\times$10.0 cm aperture area and a sandwich structure consisting of 500 $\mu$m-thick iron plates and 300 $\mu$m-thick emulsion films. (The base is 180 $\mu$m thick and the emulsion layers installed on both sides have thicknesses of 60 $\mu$m.) For additional details on the ECC, see \cite{ECC}. The multi-stage shifter is installed at the downstream side of the ECC. The shifter, which is 44.0 cm$\times$22.0 cm$\times$6.0 cm and has an aperture area of 12.5 cm$\times$10.0 cm, is composed of three stages on which doublet emulsion films are mounted. The stages are separated by 1 mm gaps. These emulsion films consist of same type of ECC's emulsion films. The power consumption is 7 W. A practical model of the multi-stage shifter was co--developed with Mitaka Kohki Co., Ltd. based on the demonstration described in \cite{MultiStageShifter} and used in the 2011 balloon experiment\cite{GRAINE2011, GRAINE2011_2}. Additional details regarding the development of the multi-stage shifter can be found in \cite{MSS_related}. In the balloon experiment, the multi-stage shifter was set horizontally and operated with the $\sim$hour scale.  In order to expose the films perpendicularly to the neutrino beam, the multi-stage shifter had to be set vertically. In addition, since the exposure period of an accelerator neutrino experiment is several months, long-term operation is required for the multi-stage shifter. To ensure these diverse requirements could be met, we tested these components in various ways for vertical driving and long--term operation prior to conducting our experiment.

The multi-stage shifter drive was started on Oct 29, 2014, and then stopped on Nov 23 to permit re-treatment of the emulsion films. On Nov 27, the multi-stage shifter was restarted and operation continued until Dec 22. After a trial-and-error period for emulsion film handling, the emulsion detector was renewed in front of the INGRID module. The drive of the multi-stage shifter was started on Jan 14, 2015. Due to beam period elongation, it was decided to modify the operation parameters from Feb 13 to elongate the operation period. Operation of the multi-stage shifter then continued to Apr 1. After beam exposure and shifter operation adjustments, in situ exposure to cosmic rays was conducted from Apr 1 to 8 in order to permit pre-alignment of the shifter films. 

\begin{figure}[h!!!!!!!!]
\centering\includegraphics[width=6.5in]{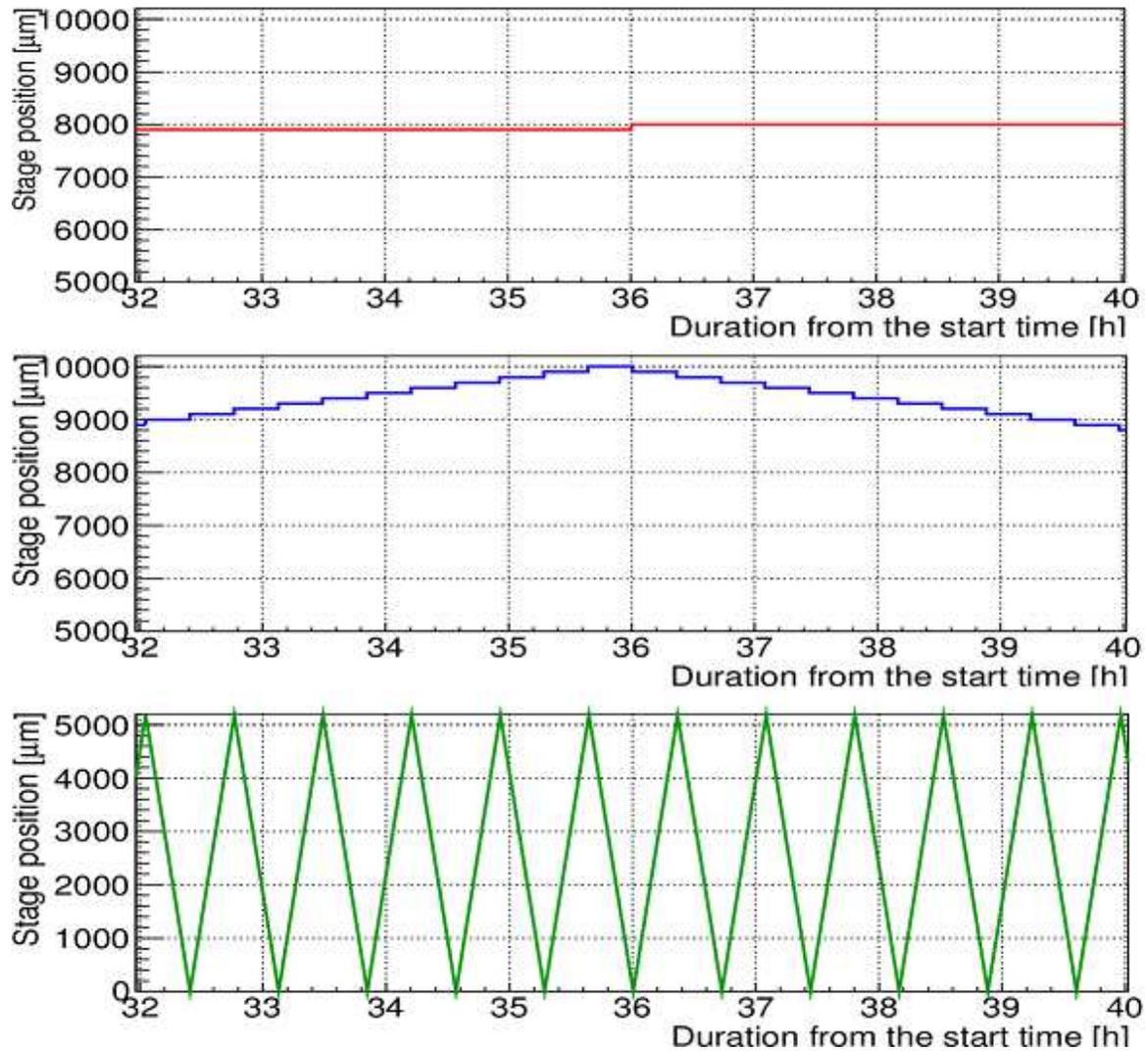}
\caption{Timing chart for portion of fourth operation (from 22:31 Feb 14 to 6:31 Feb 15, 2015). Top, middle, and bottom panels show first, second, and third stages, respectively.}
\label{operation_graph}
\end{figure}

\begin{figure}[h!!!!!!]
\centering\includegraphics[width=5.0in]{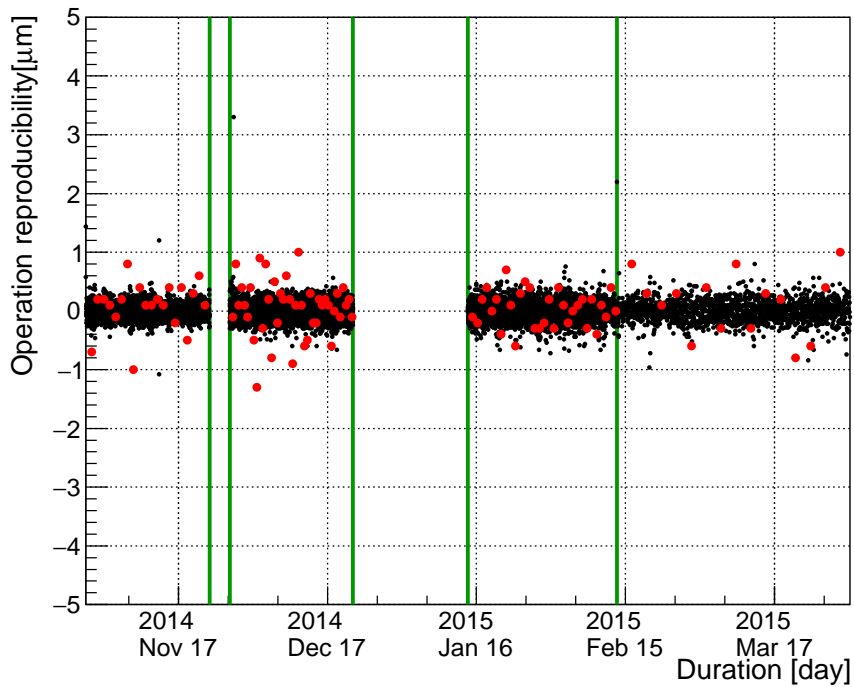}
\caption{Reproducibility of each stage during operation. Black and red plots show third and second stages, respectively.}
\label{dp_time}
\end{figure}

Table \ref{operation} shows the operation parameters for the multi-stage shifter for each operation. For example, the 20 $\mu$m/s third stage velocity in the first operation corresponds to 0.25 s time resolution with five $\mu$m connection accuracy between the second and third stage. Figure \ref{operation_graph} shows a timing chart for a portion of the fourth operation during which remote monitoring of the status of the multi-stage shifter was performed. Figure \ref{dp_time} shows the reproducibility of each stage during the operations. The stages of the multi-stage shifter were controlled by pulse motors. The third and second stages returned to the reference position for each cycle. The  reproducibility was evaluated with the difference in the number of pulses between go and return. It can be seen that root mean square (RMS) repeatabilities of 0.16 and 0.46 $\mu$m were obtained for the third and second stages, respectively, for all operation periods. More specifically, we performed stable multi-stage shifter operation for 24.9, 24.8, 30.0, and 46.9 days in continuity for each period, and achieved a total of 126.6 days of operation without encountering any problems with the multi-stage shifter. We achieved $\sim$month scale operation of the multi-stage shifter vertically.

\begin{landscape}
\begin{table}[h]
\begin{center}
\caption{Operation parameters of multi--stage shifter for each operation.}
\label{operation}
\begin{tabular}{c c|c c c|c c c c|c c c }
\shortstack{}&\shortstack{}&\multicolumn{3}{c|}{\shortstack{first stage\\(100 $\mu$m--step)}}& \multicolumn{4}{c|}{\shortstack{\\{}second stage\\(100 $\mu$m--step, 100 step--stroak,\\ 9900 $\mu$m--stroak)} }&\multicolumn{3}{c}{\shortstack{third stage\\(5184 $\mu$m--stroak)}}\\
\shortstack{period}&\shortstack{duration\\$\bigl[$day$\bigr]$}&\shortstack{interval\\$\bigl[$hour$\bigr]$}& \shortstack{\\{}\# of \\steps} &\shortstack{stroak \\$\bigl[$$\mu$m$\bigr]$} &\shortstack{interval\\$\bigl[$min$\bigr]$}& \shortstack{\\{}\# of \\steps}& \shortstack{\# of \\stroaks} &\shortstack{\# of \\cycles}& \shortstack{velocity\\$\bigl[$$\mu$m/s$\bigr]$}& \shortstack{\# of \\stroaks} &\shortstack{\# of \\cycles}\\
\hline
\shortstack{\\{}2014\\ Oct 29 16:12\\--Nov 23 14:04}&\shortstack{24.9\\{}\\{}\\{}\\{}} &\shortstack[r]{14.4\\{}\\{}\\{}\\{}}&\shortstack{42\\{}\\{}\\{}\\{}}&\shortstack{4100\\{}\\{}\\{}\\{}}&\shortstack{8.6\\{}\\{}\\{}\\{}}&\shortstack{4150\\{}\\{}\\{}\\{}}&\shortstack{41.5\\{}\\{}\\{}\\{}}&\shortstack{20.8\\{}\\{}\\{}\\{}}&\shortstack{10.0\\{}\\{}\\{}\\{}}&\shortstack{4150\\{}\\{}\\{}}&\shortstack{2075.0\\{}\\{}\\{}}\\
\shortstack{\\{}2014\\ Nov 27 16:07\\--Dec 22 10:38}&\shortstack{24.8\\{}\\{}\\{}\\{}} &\shortstack[r]{7.2\\{}\\{}\\{}\\{}}&\shortstack{83\\{}\\{}\\{}\\{}}&\shortstack{8200\\{}\\{}\\{}\\{}}&\shortstack{4.3\\{}\\{}\\{}\\{}}&\shortstack{8252\\{}\\{}\\{}\\{}}&\shortstack{82.5\\{}\\{}\\{}\\{}}&\shortstack{41.3\\{}\\{}\\{}\\{}}&\shortstack{20.0\\{}\\{}\\{}\\{}}&\shortstack{8252\\{}\\{}\\{}}&\shortstack{4126.0\\{}\\{}\\{}}\\
\shortstack{\\{}2015\\ Jan 14 14:24\\--Feb 13 13:39}&\shortstack{30.0\\{}\\{}\\{}\\{}} &\shortstack[r]{11.5\\{}\\{}\\{}\\{}}&\shortstack{63\\{}\\{}\\{}\\{}}&\shortstack{6200\\{}\\{}\\{}\\{}}&\shortstack{6.9\\{}\\{}\\{}\\{}}&\shortstack{6240\\{}\\{}\\{}\\{}}&\shortstack{62.4\\{}\\{}\\{}\\{}}&\shortstack{31.2\\{}\\{}\\{}\\{}}&\shortstack{12.5\\{}\\{}\\{}\\{}}&\shortstack{6240\\{}\\{}\\{}}&\shortstack{3120.0\\{}\\{}\\{}}\\
\shortstack{\\{}2015\\ Feb 13 14:31\\--Apr 1 12:54}&\shortstack{46.9\\{}\\{}\\{}\\{}} &\shortstack[r]{36.0\\{}\\{}\\{}\\{}}&\shortstack{32\\{}\\{}\\{}\\{}}&\shortstack{3100\\{}\\{}\\{}\\{}}&\shortstack{21.6\\{}\\{}\\{}\\{}}&\shortstack{3127\\{}\\{}\\{}\\{}}&\shortstack{31.3\\{}\\{}\\{}\\{}}&\shortstack{15.7\\{}\\{}\\{}\\{}}&\shortstack{4.0\\{}\\{}\\{}\\{}}&\shortstack{3127\\{}\\{}\\{}}&\shortstack{1563.5\\{}\\{}\\{}}\\

\end{tabular}
\end{center}
\end{table}

\end{landscape}

\newpage

\section{Analysis}

\subsection{Scanning and track reconstruction}
The emulsion films exposed from Jan 14 to Apr 8, 2015 were recovered and developed. The emulsion films recorded cosmic ray muons (majority) and tracks from neutrino interactions in the wall and the ECC. The emulsion films were scanned using the fully automated hyper track selector (HTS) scanning system at Nagoya University, Japan. Through this scanning, the recognized track information (which consisted of the positions, angles, and darkness of the tracks) was obtained. The scan area and accepted angular range were 105 mm$\times$80 mm (10 mm inside from edge) and $|\tan{\theta}_{x}| \leq$ 1.8 ($|\theta_{x}| \leq$ 60.9 degree), respectively, where $\theta_{x}$ is the projection angle on the x--z plane for the normal direction of the x--axis as well as $\theta_{y}$.

The scanned tracks were connected in the following order: the most downstream of the ECC film $\rightarrow$ first stage $\rightarrow$ second stage $\rightarrow$ third stage. The tracks between films were reconstructed by using the position and angle coincidence of tracks that were extrapolated to the center of the gap between stages. In this analysis, we focused on $|\tan{\theta}_{x}| \leq$ 1.0 ($|\theta_{x}| \leq$ 45.0 degree), $|\tan{\theta}_{y}| \leq$ 1.0 ($|\theta_{y}| \leq$ 45.0 degree) angle range. Using track reconstruction, track displacement between the films on each stage was obtained. The displacement of tracks reconstructed between the films on each stage corresponds to the stage position with each timing. Using the combination of each stage position, track incident time was obtained. In this analysis, we focused on the incident tracks for the period from Feb 13 to Apr 1, 2015 because the emulsion film condition is better for the period with closer to the development date, and rather small fading effects shown at section \ref{fading}.

\subsection{Neutrino beam signature}
\begin{figure}[!h]
\centering\includegraphics[width=5.0in]{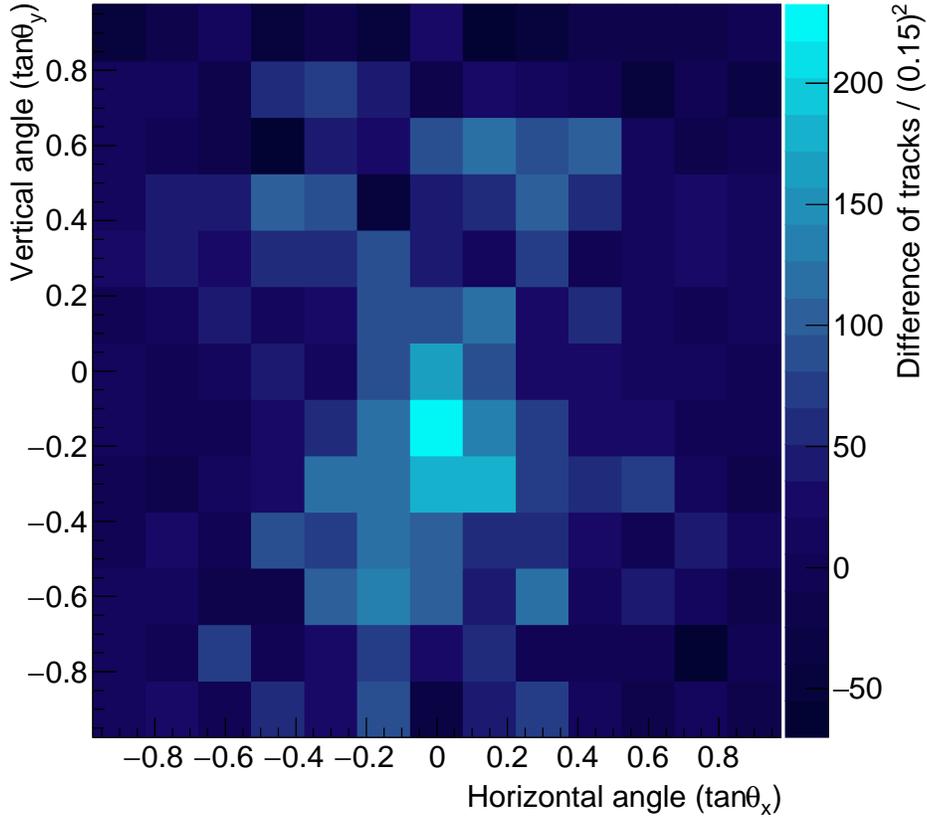}
\caption{Difference in track angle distribution between the beam--on and beam--off periods (normalized by the period and to be zero with $|\tan{\theta_y}|\geq$0.7). In vertical angle, positive indicates upwards and negative indicates downwards in the direction from the upstream to downstream of the beam (specifically, a positive direction of the z--axis).}
\label{beam}
\end{figure}

From the first stage position (divided every 36 hours), the beam-on period and beam-off period (pre-alignment period) can be distinguished. Figure \ref{beam} shows the difference in the track angle distribution between the beam--on and beam--off periods. The beam excess (so-called sand muons which are induced at a wall in front of the detector by the muon neutrino beam) was clearly detected at a center in the horizontal angle and at a slightly downward direction in the vertical angle due to the beam with a slightly downward direction. We then used time information to find the neutrino beam signature.

\subsection{Measurement of fading effect \label{fading}}

\begin{figure}[!h]
\centering\includegraphics[width=5.0in]{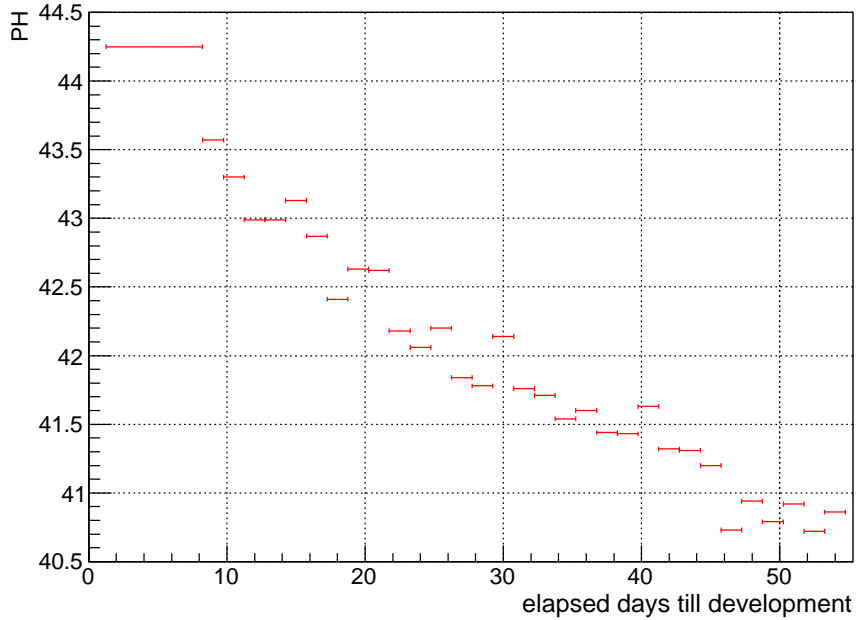}
\caption{Pulse Height (PH), defined as the sum of four emulsion layers (most downstream of the ECC film and the first stage film) peak values, as a function of elapsed time. The first point indicates the pre-alignment period. The other points indicate 36 h periods. The emulsion films were kept refrigerated from recovery to development.}
\label{ph}
\end{figure}

The darkness of the emulsion tracks correlates with $dE/dx$ for the incident particles and can be used for particle identification. The track darkness information is obtained during track recognition, and is referred to as the pulse height (PH). When the scan system recognizes the tracks, 16 tomographic images in an emulsion layer are obtained. PH is defined as the number of hit images. Figure \ref{ph} shows PH [the sum of four emulsion layers (most downstream of the ECC film and the first stage film) peak value] as a function of the elapsed time until development. Here, PH attenuation was clearly seen (8\%). This can be explained as latent image fading. We, for the first time, used time information to measure precise fading effects using tracks recorded in the same emulsion film. These correspond to quite same conditions for film characteristics, incident particles, elapsed environment, development effects, and scanning conditions. By measuring the fading effects, the PH can be corrected even if the tracks degrade. Thus, it also provides us with precise $dE/dx$ measurements. In this exposure, the emulsion detector was maintained at a temperature of about 22 ${}^\circ\mathrm{C}$. It is possible to suppress track fading by maintaining the emulsion films at a low temperature (10 ${}^\circ\mathrm{C}$ or below).

\subsection{Measurement of track rate}
\begin{figure}[!h]
\centering\includegraphics[width=5.0in]{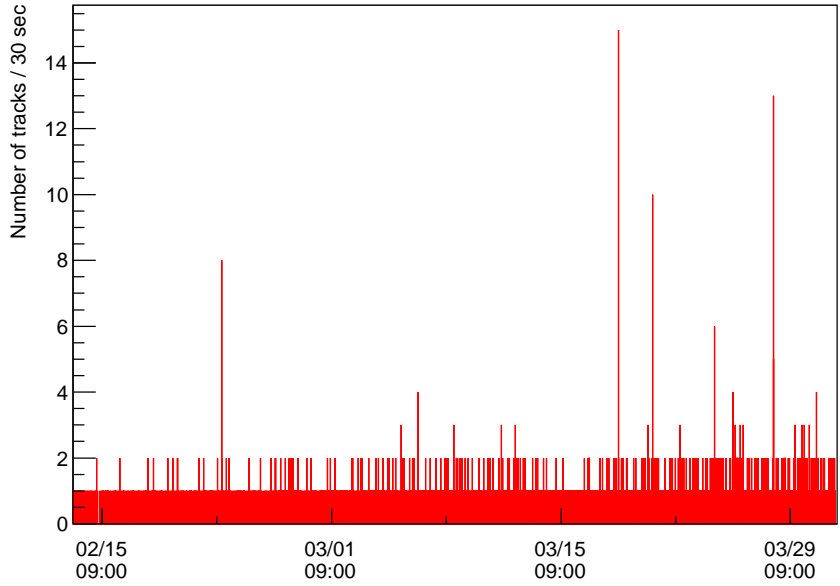}
\caption{Number of tracks per 30s as a function of operation period.}
\label{track_rate}
\end{figure}

\begin{figure}[!h]
\centering\includegraphics[width=5.0in]{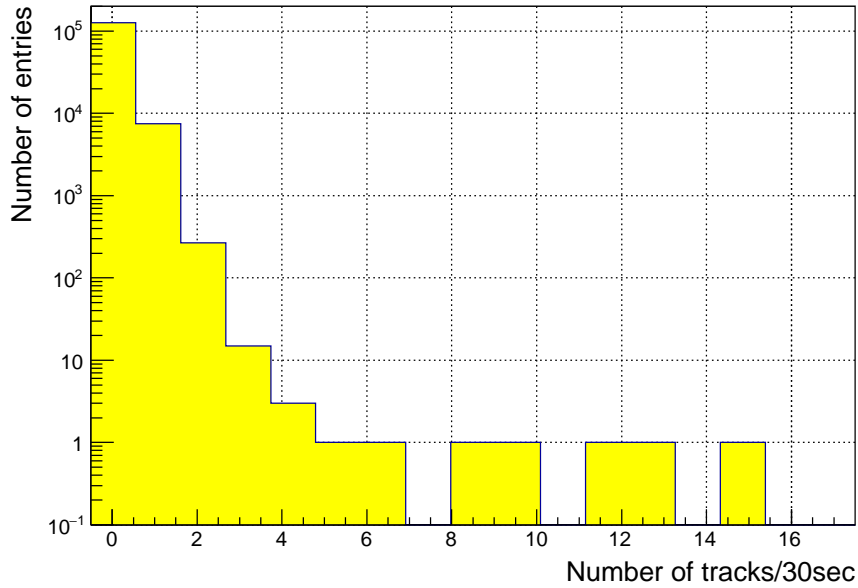}
\caption{Histogram showing number of tracks per 30s.}
\label{count_hist}
\end{figure}

From the third stage position, all incident tracks had time information at a resolution of $\sim$seconds.  Figure \ref{track_rate} shows the number of tracks per 30 s as a function of the operation period. The number of tracks was basically constant during operation. For earlier periods, the number of tracks appears to decrease slightly due to fading effects. In addition, spikes appear on a number of tracks. Figure \ref{count_hist} shows the histogram of the number of tracks per 30 s. 

\newpage

\subsubsection{Detection of high multiplicity events}
\begin{figure}[!h]
\centering\includegraphics[width=5.0in]{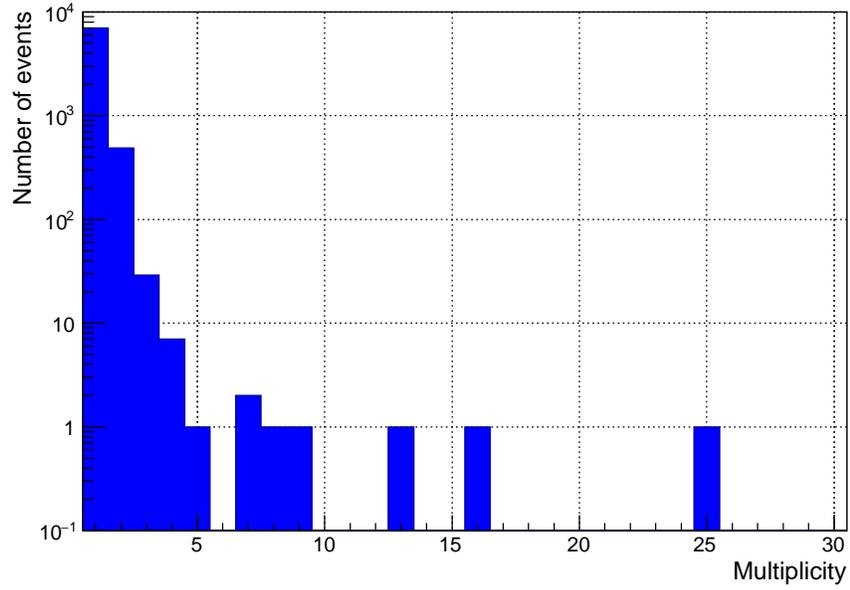}
\caption{Multiplicity distribution after clustering.}
\label{clster_hist}
\end{figure}

\begin{figure}[!h]
\centering\includegraphics[width=5.0in]{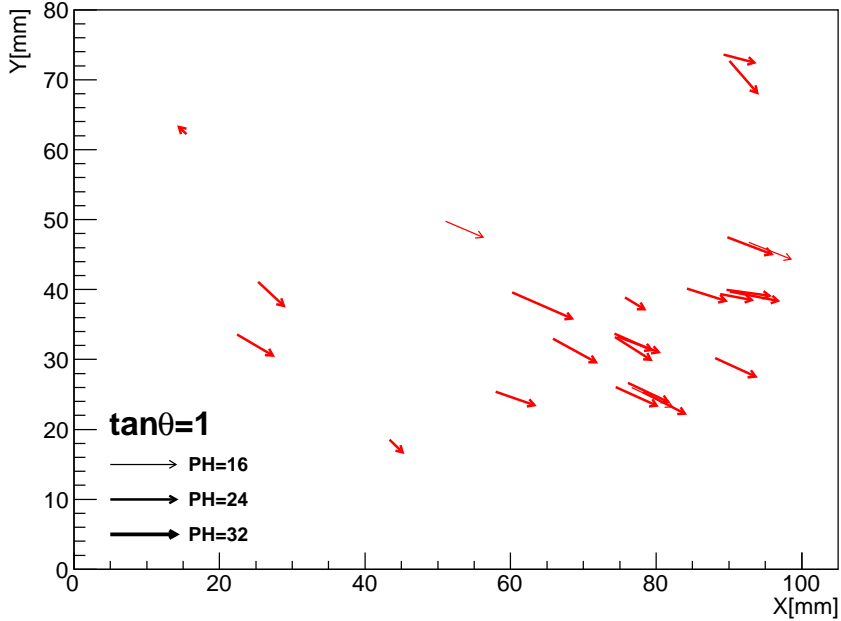}
\caption{Vector map of high multiplicity event (25 tracks).
Horizontal and vertical ranges show the scanned area of the most downstream film of the ECC. The starting point of the vectors indicates the position of the tracks on the same film. The length and direction of the vectors indicate the size and direction of the incident angle of the tracks. The end point of the vectors indicates the position of the tracks extrapolated onto the third stage film of the multi-stage shifter. The vector widths indicate of the PH [sum of two emulsion layers (most downstream of the ECC film)] of the tracks. The vectors (black) located at the bottom-left of the vector map indicate the angle scale and PH scale.}
\label{multi_event}

\end{figure}

We found some excesses in the number of tracks per 30 s (Figure \ref{count_hist}). Tracks localized in time were clustered. The clustered events had no tracks beyond 30 s from the timing edge of the event. Figure \ref{clster_hist} shows the multiplicity distribution after clustering. We found seven clear high multiplicity events ($\geq$ seven tracks) : two events with seven tracks, single events with eight, nine, 13, 16, and 25 tracks. Figure \ref{multi_event} shows a vector map of the one of high multiplicity events (with 25 tracks). These tracks had similar incident angles, as did the other events. In addition, no events that matched with INGRID were found. It was expected that the events occurred during off-timing of the beam. Therefore, we assumed that the events were induced by cosmic rays.

\newpage

\subsubsection{Evaluation of the time resolution\label{sec_time_res}}
\begin{figure}[!h]
\centering\includegraphics[width=5.0in]{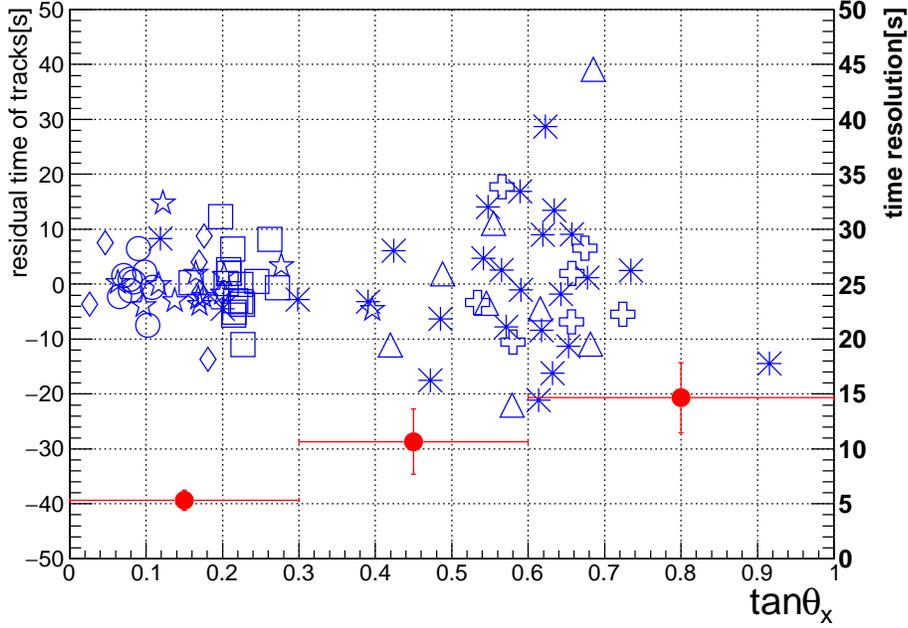}
\caption{Plots with different symbols without error bars indicate the residual time (left vertical axis) for tracks for each high multiplicity event. Plots with filled circles and error bars indicate the standard deviation (right vertical axis) of the residual time for each incident angle.}
\label{time_resolution}
\end{figure}

Tracks in the high multiplicity events were expected to be recorded simultaneously. The time resolution of the multi-stage shifter was evaluated by using seven high multiplicity events.  For each event, an average incident time was calculated and the residual time between the average incident time and the incident time for the individual tracks was obtained. Figure \ref{time_resolution} shows the residual time for tracks for each high multiplicity event. From these, the standard deviation of the residual time, which was defined as the time resolution, was obtained. Figure \ref{time_resolution} shows the standard deviation of the residual time for each incident angle. The evaluated time resolution of the multi-stage shifter in this operation was 5.3--14.7 s. We evaluated the time resolution by using the previously mentioned high multiplicity events and performed an evaluation up to the large incident angle used in this analysis. In this operation, the third stage velocity was reduced in order to elongate the operation period. By increasing the velocity, the time resolution can be improved. Moreover, by decreasing the gap between stages and increasing the number of stages, the time resolution can be improved by several orders of magnitude. This allows beam--on/off timing ($\sim$2 s cycle) to be distinguished.

\subsubsection{Reconstruction of vertex events\label{reconstruction}}

\begin{figure}[!h]
\centering\includegraphics[width=4.5in]{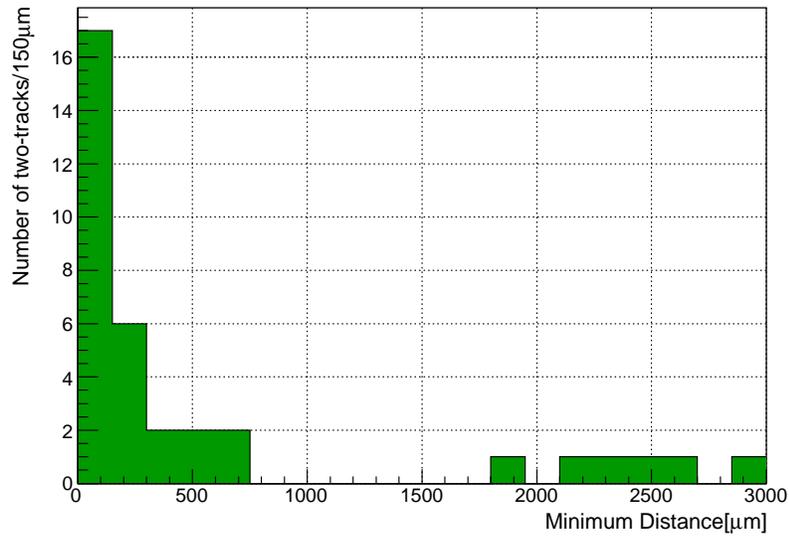}
\caption{Distribution of minimum distances.}
\label{md}
\end{figure}

\begin{figure}[!h]
\centering\includegraphics[width=4.5in]{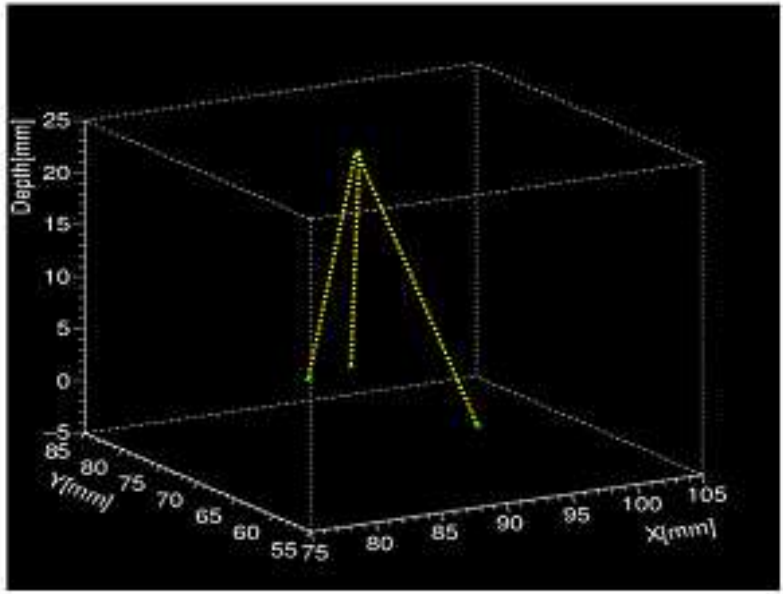}
\caption{One of the vertex events. Solid lines (green) show track segments in the most downstream of the ECC film. Dashed lines (yellow) show the lines extrapolated to the vertex point. Tracks labeled ''1'' and ''2'' were confirmed to be charged particles and the other track labeled ''e-pair'' was confirmed to be an electron pair in the ECC.}
\label{neu_event}
\end{figure}

\begin{figure}[!!!!!!!!!!h]
\centering\includegraphics[width=4.5in]{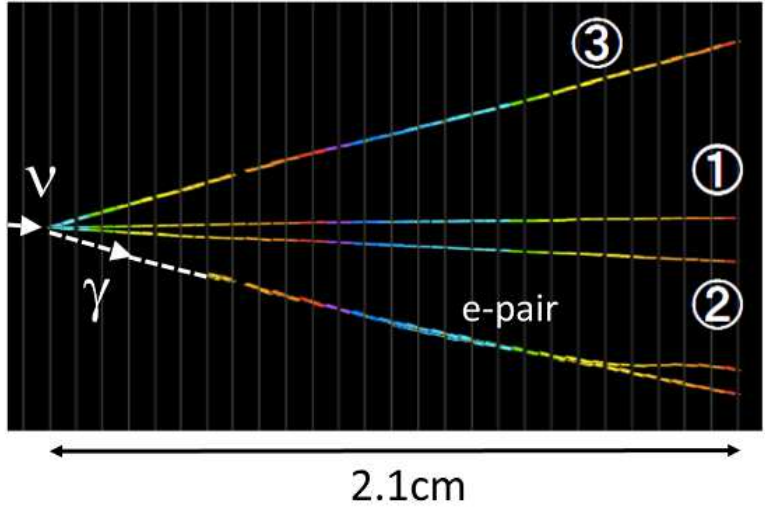}
\caption{An ECC event display. Tracks labeled ''1'', ''2'' and ''3''  were charged particles and the other track labeled ''e-pair'' was an electron pair.}
\label{nu_ECC_display}
\end{figure}

Next, we searched clustered events for vertex events with low multiplicity (2--5 tracks). To accomplish this, the minimum distance between any two tracks in a clustered event was calculated. Figure \ref{md} shows a distribution of the minimum distances. We found events localized around small minimum distances. Here, 29 track pairs with minimum distances of less than 1 mm were selected. Chance coincidence was estimated to be less than 2.0 track pairs from the event density for large minimum distances (1--3 mm). By improving the time resolution, we found that chance coincidence could be reduced. After track pair clustering for 29 track pairs, 23 clustered events (vertex events) were reconstructed. Of these 23 vertex events, 20 events had two tracks and three events had three tracks. A major component of the 20 events with two tracks was an electron pair candidate which has a topology with close position and narrow angle between the two tracks due to a small opening angle of an electron pair. Some of the events were confirmed to be electron pairs by checking the two tracks starting from same position in the ECC. Figure \ref{neu_event} shows a vertex event that was confirmed to be a neutrino interaction by checking the tracks starting from same position in the ECC. Figure \ref{nu_ECC_display} shows an ECC display of that event. We performed a reconstruction of the neutrino interaction with timing and spatial information by using the multi-stage shifter. In addition, we detected a $\gamma$-ray attaching to  the neutrino interaction. These results show that inclusive event reconstructions, including $\gamma$-ray and neutral particle events, can be performed. Moreover, by detecting a event with two $\gamma$-rays using timing and spatial information, the 1--$\pi^{0}$ production event could also be reconstructed.


\newpage
\clearpage

\subsection{Matching of emulsion and INGRID events}
\subsubsection{Sand muon matching\label{sand muon matching}}
Next, using emulsion tracks time stamped to the third stage, we attempted to perform event matching to INGRID events. The efficiency of time stamped application to emulsion tracks was $\sim$80\% in the track base obtained from neutrino event timestamping as shown in section \ref{nu_eve_matching}. (The timestamp inefficiency was consistent with the fading effect. Since two emulsion films are mounted on each stage, taking both into consideration permits the timestamp efficiency to be recovered. Moreover, by suppressing the fading effect, the timestamp inefficiency can be reduced.) 

\begin{figure}[!!!!!!!!!!h]
\centering\includegraphics[width=5.0in]{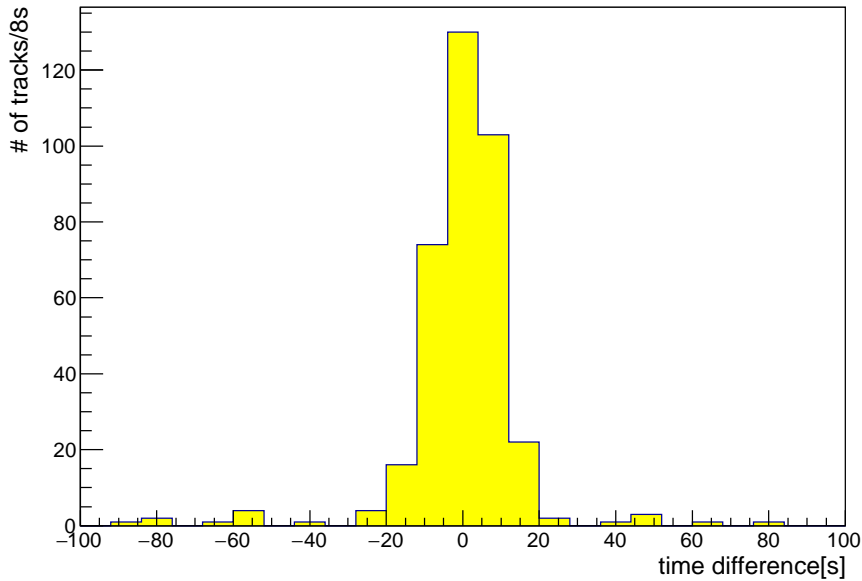}
\caption{Time difference between emulsion track and INGRID events.}
\label{sandmuon}
\end{figure}

The following criteria were required for selection as INGRID events: use of beam--on timing; having, at least, one track with three hits or more on each projection; having a hit on the most upstream plane. The INGRID events were mainly sand muons, and the hit efficiency for sand muons was 98\%. In this analysis, averaged INGRID event rate was 14.1 events per 200 s. INGRID events were selected with a $\pm$ 100 s time window for the timestamp of each emulsion track. Of these, events with position and angle differences within 3$\sigma$ were used (In this analysis, obtained position difference $\sigma$s were 1.7 cm and 1.7 cm for each projection, respectively. And angle difference $\sigma$s were 0.057 rad and 0.047 rad for each projection, respectively. These $\sigma$s were dominated by a track position and angle accuracy of the INGRID.). Figure \ref{sandmuon} shows the time difference between the emulsion track and the INGRID event. As can be seen in the figure, we obtained a clear matching peak and a fitted $\sigma$ of 7.9 s, which corresponds well to the absolute timing resolution of the multi-stage shifter for 46.9 days in this operation. It is also consistent with the time resolution evaluated at section \ref{sec_time_res}. In this operation, in order to elongate the operation period, the third stage velocity was slowed. By making fast, the time resolution can be improved. Moreover, by decreasing the gaps between stages and by increasing the number of stages, the time resolution can be even more significantly improved. By setting the time difference within 3$\sigma$ and using a chance coincidence density above 3$\sigma$, a matching reliability 98.5\% was obtained. Based on the above, we can confirm that emulsion track matching to INGRID events with high reliability was performed using the multi-stage shifter.

\newpage

\subsubsection{Neutrino event matching \label{nu_eve_matching}}

We detected neutrino events (three events and five tracks) with the ECC and followed them down to the most downstream of the ECC films. Of these five tracks, four tracks in three events were timestamped (One track showed a timestamp inefficiency that was consistent with the fading effect. The timestamp efficiency can be recovered and improved as shown at section \ref{sand muon matching}.). In addition, we detected neutrino events by reconstructing vertex events with the multi-stage shifter, as shown in section \ref{reconstruction}, and confirmed neutrino events in the ECC (three events and eight tracks). By combining the neutrino events detected with the ECC and the multi-stage shifter, we confirmed four events and nine tracks\footnote{Two events and three tracks, respectively, were overlapped.} (Two events with three tracks, one event with two tracks\footnote{Not an electron pair due to more than two tracks from the primary vertex}, and one single track event). Track timestamps in each event were consistent with the timing resolution. We timestamped neutrino events using the average track timestamps, and then attempted neutrino event matching with the INGRID. INGRID events were selected using the event timing differences, track positions, and angles within 3$\sigma$. (These are the same criteria used for sand muon matching shown in section \ref{sand muon matching}). 

\begin{figure}[!h]
\centering\includegraphics[width=5.5in]{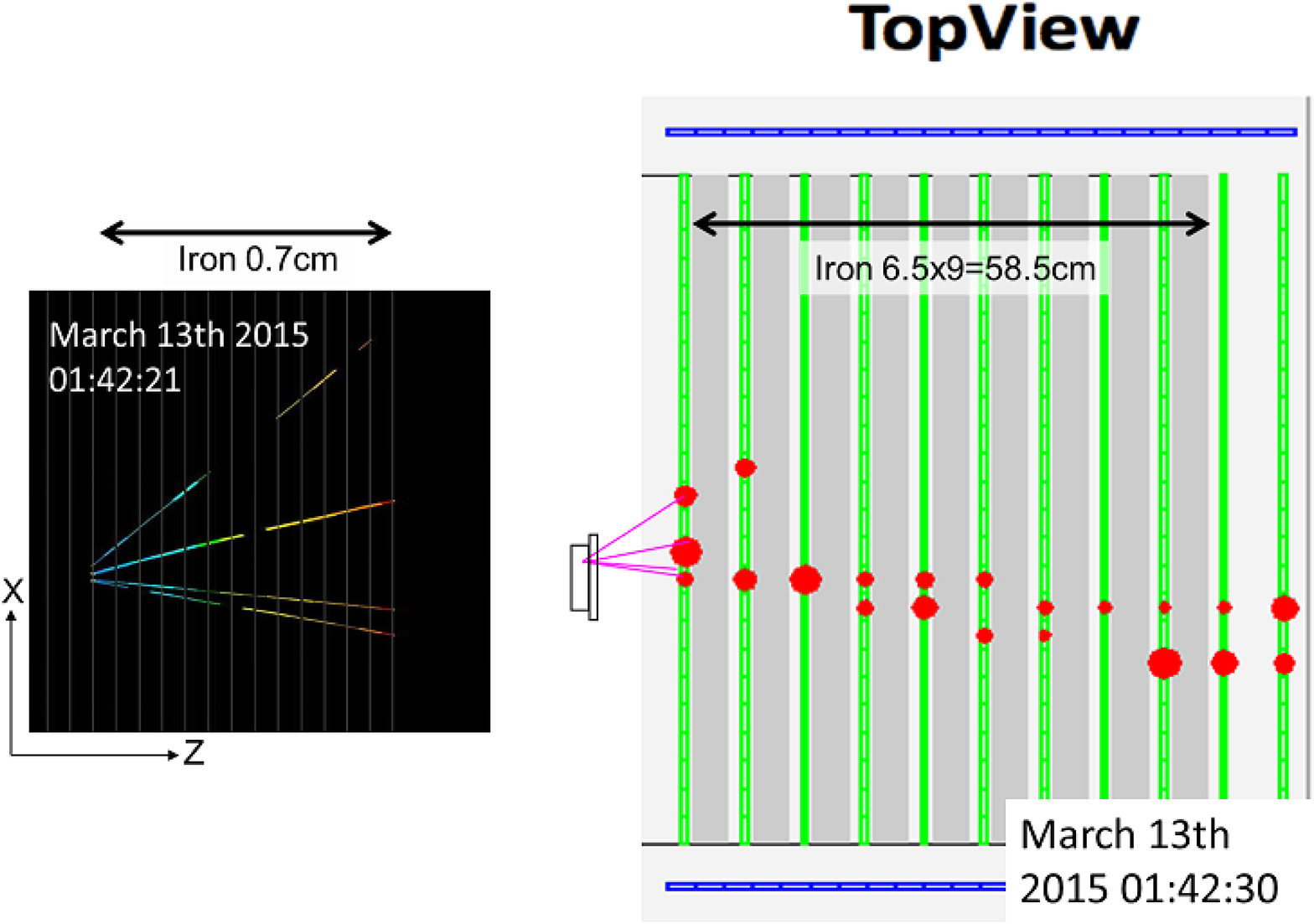}
\centering\includegraphics[width=5.5in]{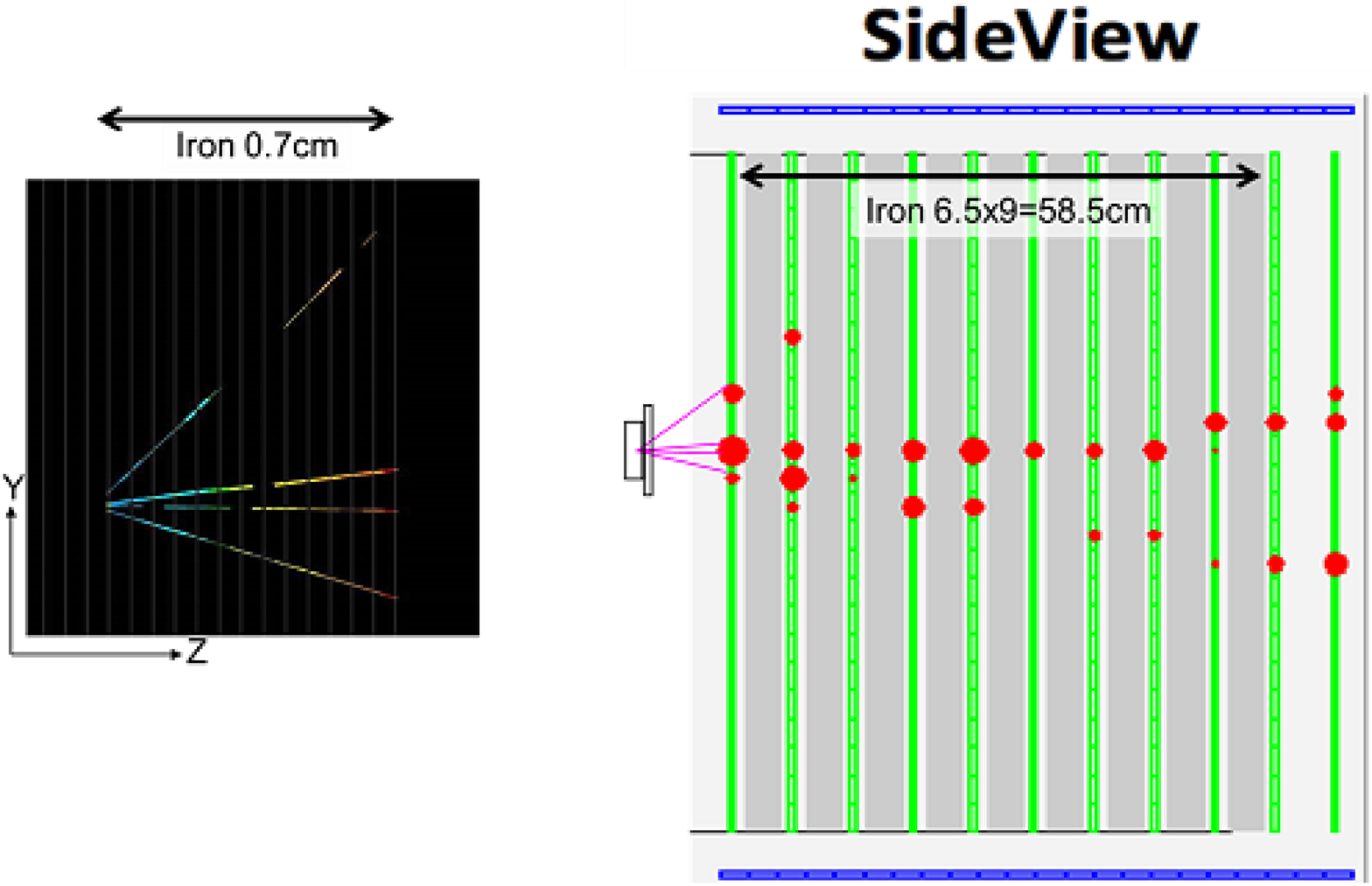}
\caption{Matched event \#1. Event Displays of ECC (left) and INGRID (right). Top and bottom panel show top and side view, respectively. } 
\label{nu_first}
\end{figure}

\begin{figure}[!h]
\centering\includegraphics[width=5.5in]{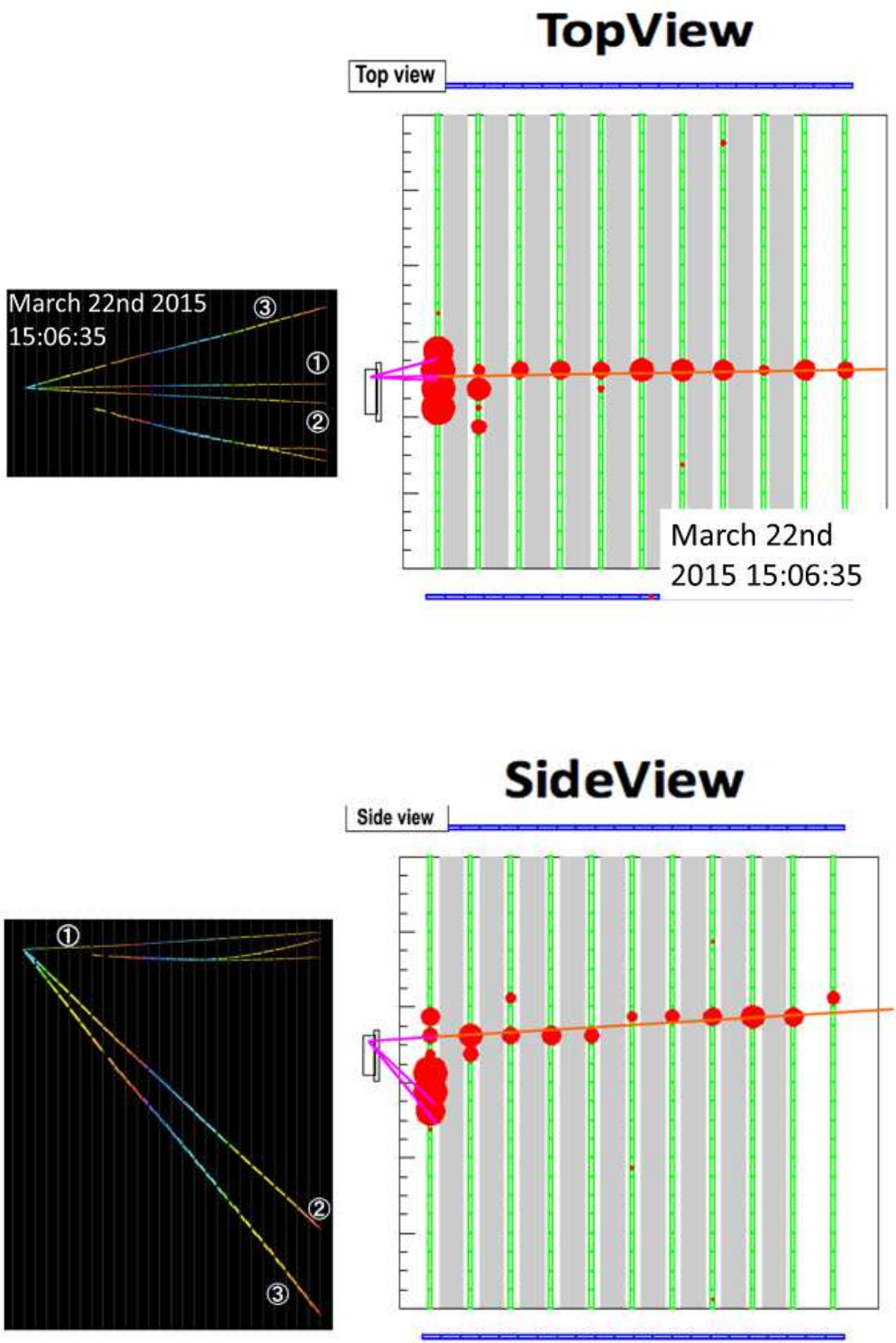}
\caption{Matched event \#2.} 
\label{nu_second}
\end{figure}

\begin{figure}[!h]
\centering\includegraphics[width=4.75in]{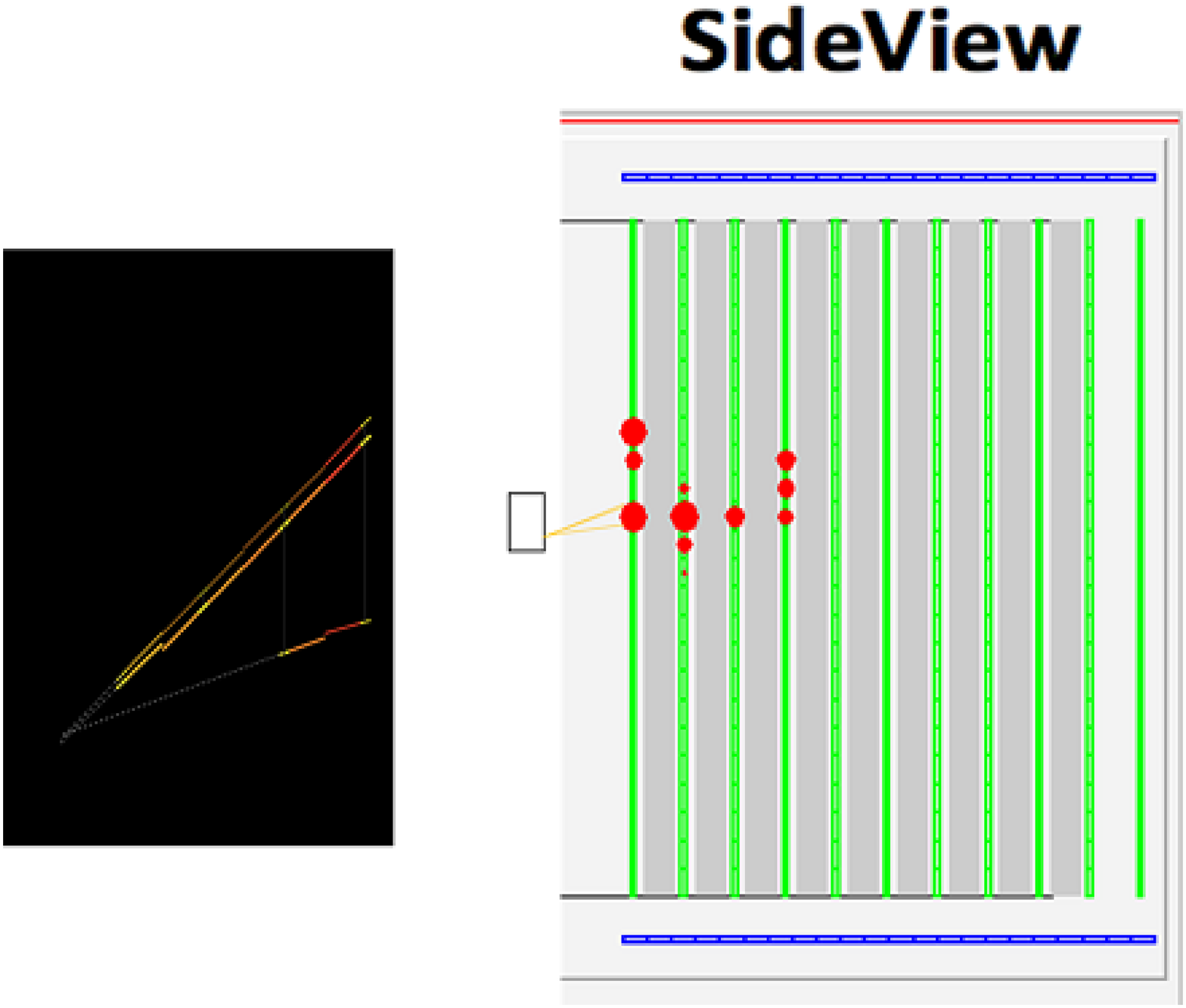}
\caption{Matched event \#3.} 
\label{nu_third}
\end{figure}

Of the four events, three were uniquely matched with INGRID events. The matched events had reasonable timing differences. Figure \ref{nu_first}, \ref{nu_second}, and \ref{nu_third} show ECC and INGRID event displays. In addition, the matched events were topologically and kinematically compatible. Based on the above, we succeeded in event matching with INGRID and established a hybrid analysis by using the multi-stage shifter. The single event with one track was not matched with an INGRID event. The track in this event was black track and not a minimum ionization particle (to a high $dE/dx$ particle). Using measurements of multiple Coulomb scattering, $p\beta$ was 22 MeV/$c$ at the multi-stage shifter. Here, $p$ and $\beta$ are the momentum and the velocity in units of light speed, $c$, which corresponds to a kinetic energy value less than 12 MeV. The energy loss in a 6.5 cm thick steel plate with a plane in the INGRID module is 74 MeV for the minimum ionization particle. Since INGRID events are required to have tracks with more than three-plane hits and thus, it is likely that this track is not identified as an INGRID event.
\newpage

\newpage

\begin{figure}[!h]
\centering\includegraphics[width=5.5in]{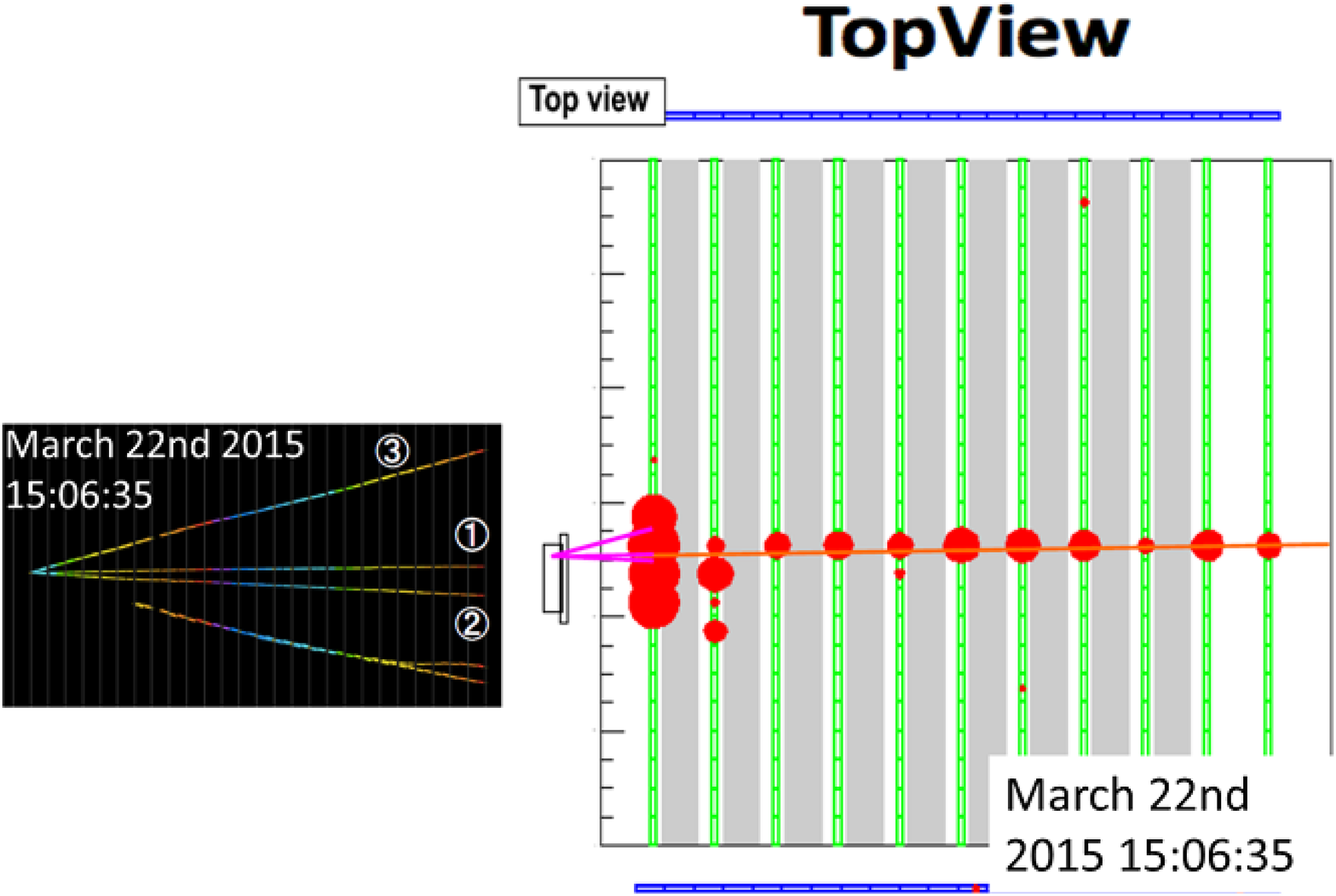}
\end{figure}

\newpage

\begin{figure}[!h]
\centering\includegraphics[width=4.75in]{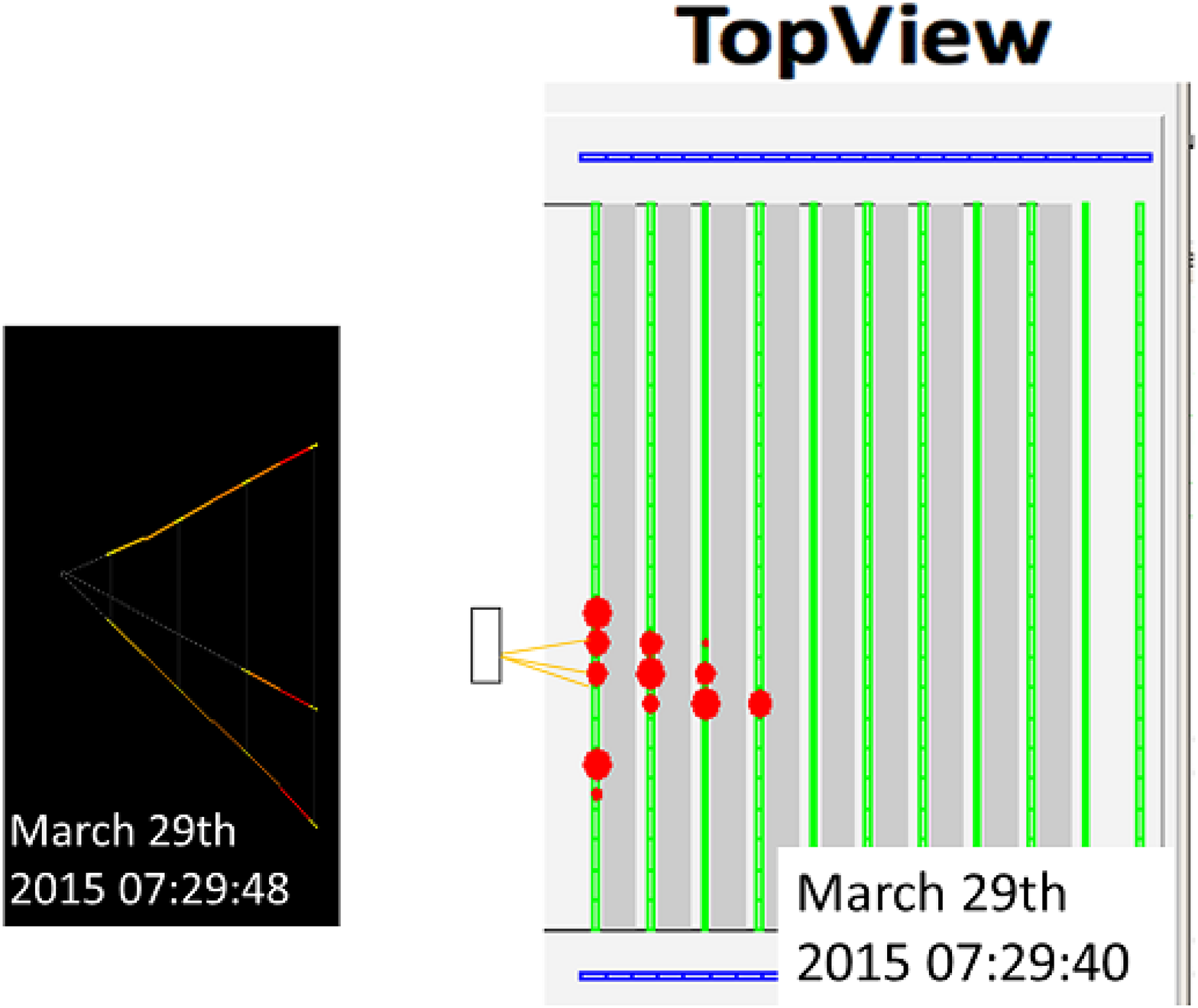}
\end{figure}

\newpage

\section{Summary}
In this paper, we reported on J-PARC T60 experiments using an emulsion detector for the feasibility studies on precise measurement of $\nu$--N interactions and strict validation of neutrino oscillation anomalies. In our accelerator neutrino experiments, a multi-stage shifter was implemented with vertical and month scale operation for the first time, and stable operation (0.16 and 0.46 $\mu$m reproducibility of third and second stages) was performed for a total of 126.6 days. 

The emulsion films exposed from Jan 14 to Apr 8, 2015 were scanned ($|\theta_{x,\ y}| \leq$ 60.9 degree), and operation over the period of Feb 13 to Apr 1, 2015 was analyzed ($|\theta_{x,\ y}| \leq$ 45.0 degree). 

Using the first stage position, the neutrino beam signature was clearly detected. In addition, fading effects were measured for the first time using tracks recorded in the same emulsion film. These correspond to quite the same  conditions for film characteristics, incident particles, elapsed environment, development effects, and scanning conditions. By precisely measuring fading effects, the PH could be corrected even if tracks exhibited such effects. The results provide us with $dE/dx$ measurements.

From the track rate measured using the third stage position, high multiplicity events were clearly identified. Using these high multiplicity events, a time resolution of 5.3 to 14.7 s was evaluated over a 46.9--day operation period (4.05$\times10^{6}$ s). This result corresponds to time resolved numbers of (3.8 to 1.4)$\times10^{5}$ ($\frac{4.05 \times 10^{6}}{(5.3\ {\rm to}\ 14.7) \times 2}$). In this operation, to elongate the operation period due to beam period elongation, the third stage velocity had to be reduced. By increasing the velocity, the time resolution can be improved. Moreover, by increasing a number of stages and decreasing the gap between them, we found that the time resolution could be improved by several orders of magnitude. As a result, beam--on/off timing ($\sim$2 s cycle) could be distinguished.

A reconstruction of vertex events was performed using timing and spatial information, from which neutrino events were identified. In addition, a $\gamma$-ray event attaching to the neutrino event was found. Thus, it was clarified that event reconstruction that included $\gamma$-ray and neutral particle events could be performed, and that a 1--$\pi^{0}$ production event could be reconstructed.

Emulsion track matching to INGRID events was performed with high reliability (98.5\%). By using the matching peak, we obtained an absolute time resolution of 7.9 s, which is consistent with the time resolution evaluated by using the high multiplicity events. In addition, neutrino events were timestamped and clear neutrino event matching with the INGRID events was performed. Furthermore, hybrid analysis was established by using the multi-stage shifter. 

Finally, from the results shown in this paper, it is clear that we have, for the first time, demonstrated the feasibility of multi-stage shifter use with accelerator neutrino experiments.

\section*{Acknowledgment}
The authors would like to express their appreciation for the support provided by the J-PARC and the T2K Collaboration. This work was supported by the Japan Society for the Promotion of Science (JSPS) KAKENHI (Grant Numbers 20244031, 25105001, 25105006, 25707019, 26105516, 26287049, 26800138, 16H00873).


%

\appendix

\end{document}